\documentclass[prl,twocolumn,a4paper]{revtex4}
\usepackage{graphicx}
\usepackage{bm}
\usepackage[T1]{fontenc}
\usepackage{amsmath,amssymb}
\usepackage{mathrsfs} 
\usepackage{times}
\usepackage{ulem}
\usepackage{color}
\usepackage{textcomp}

\begin{document}

\title{Remote scanning for ultra-large field of view in wide-field microscopy and full-field OCT}

\author{Ga\"elle Recher}
\affiliation{LP2N, Laboratoire Photonique Num\'erique et Nanosciences, Univ. Bordeaux, F-33400 Talence, France}
\affiliation{Institut d\textquotesingle Optique Graduate School $\&$ CNRS UMR 5298, F-33400 Talence, France}
\author{Pierre Nassoy}
\affiliation{LP2N, Laboratoire Photonique Num\'erique et Nanosciences, Univ. Bordeaux, F-33400 Talence, France}
\affiliation{Institut d\textquotesingle Optique Graduate School $\&$ CNRS UMR 5298, F-33400 Talence, France}
\author{Amaury Badon*}
\affiliation{LP2N, Laboratoire Photonique Num\'erique et Nanosciences, Univ. Bordeaux, F-33400 Talence, France}
\affiliation{Institut d\textquotesingle Optique Graduate School $\&$ CNRS UMR 5298, F-33400 Talence, France}

\email{amaury.badon@institutoptique.fr}

\date{\today}

\begin{abstract}

Imaging specimens over large scales and with a sub-micron resolution is instrumental to biomedical research. Yet, the number of pixels to form such an image usually exceeds the number of pixels provided by conventional cameras. While most microscopes are equipped with a motorized stage to displace the specimen and acquire the image tile-by-tile, we propose an alternative strategy that does not require any moving part in the sample plane. We propose to add a scanning mechanism in the detection unit of the microscope to collect sequentially different sub-areas of the field of view. Our approach, called remote scanning, is compatible with all camera-based microscopes. We evaluate the performances in both wide-field microscopy and full-field optical coherence tomography and we show that a field of view of 2.2 mm with 1.1 $\mu$m resolution can be achieved. We finally demonstrate that the method is especially suited to image biological samples such as millimetric engineered tissues.

\end{abstract}

\maketitle

\section{Introduction}

Deciphering biological processes often relies on the dual ability to perform in toto three-dimensional imaging of the sample, which can be of macroscopic size, and  to  reach a sub-cellular (i.e.  $\sim$ $\mu$m) resolution. Amongst all available biomedical imaging techniques, optical microscopy, which is characterized by a high spatial resolution and a moderate invasiveness, is undoubtedly a good candidate. In the recent years, most efforts have focused on developing strategies to image living samples in depth. However, imaging large volumes with sub-micron resolution also raises an issue that has been widely overlooked, namely the limited amount of information that can be captured by a microscope. 

\begin{figure}
    \centering
    \includegraphics[width=0.45\textwidth]{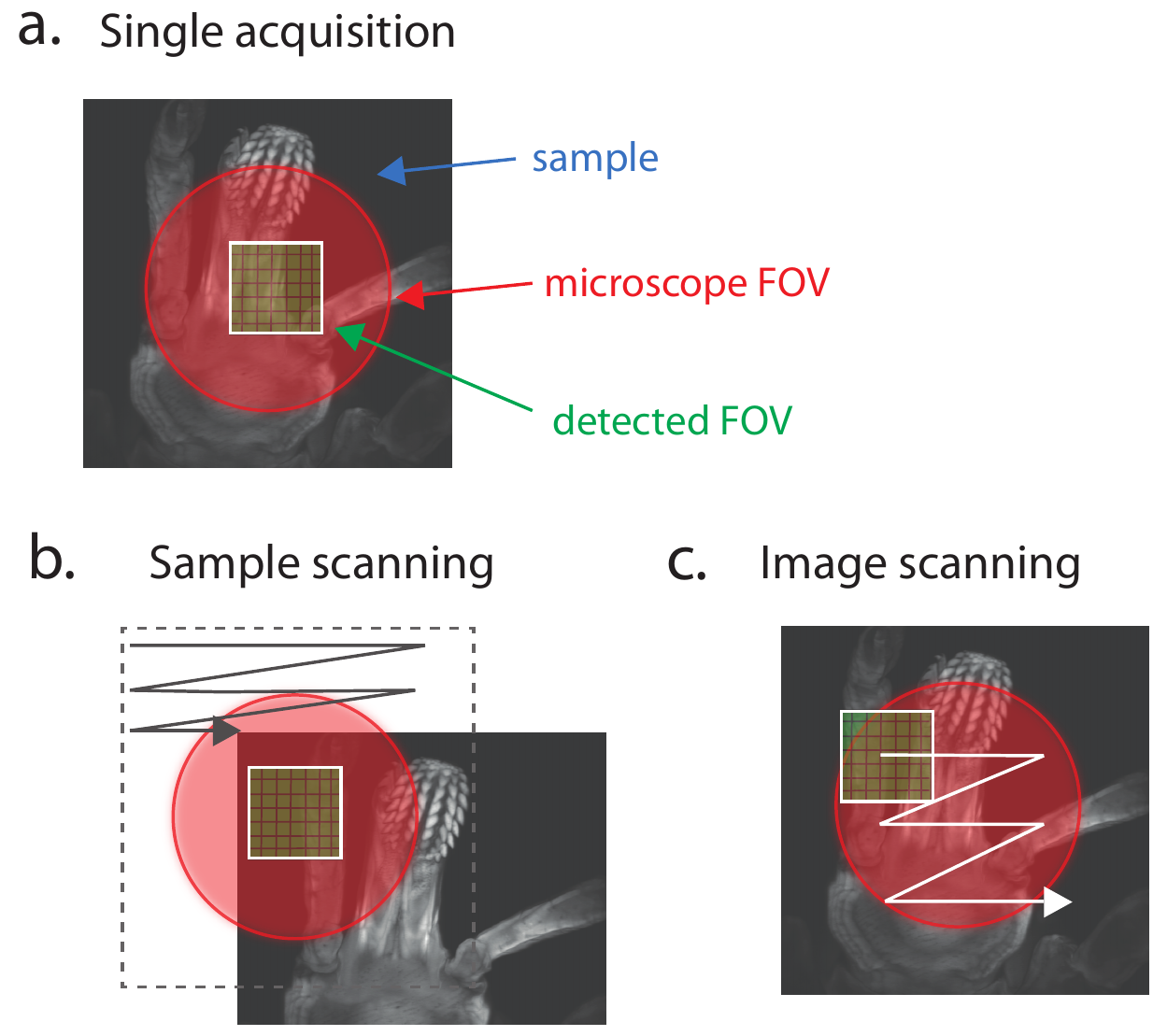}
    \caption{Principle of the remote scanning approach. (a) With a conventional microscope, the recorded FOV is limited by the detected area which is usually much smaller than the accessible FOV of the MO. (b) The simplest solution to enlarge the FOV consists in moving the specimen and keeping the detection still. (c) Alternatively, access to a larger FOV can be obtained by scanning the detected area across the FOV accessible by the MO, the specimen remaining stationary.}
    \label{fig1}
\end{figure}

It is common to consider that the microscope objective (MO) is the primary limiting factor of a microscope throughput. Indeed, we all experienced that a high magnification MO provides a small field of view (FOV) at high lateral resolution, while a low magnification MO would expand the FOV but with a reduced resolution. Quantitatively, the spatial bandwidth product (SBP) characterizes the throughput of an objective. It is defined as the number of pixels necessary to capture the full FOV at Nyquist sampling \cite{lohmann1996space,lukosz1966optical,mcconnell2019video}. As a reference point, a standard X10 with a resolution 1.1 $\mu$m and a 2.65 $\times$ 2.65 $\text{mm}^2$ FOV has an SBP of 21 megapixels. This number already exceeds the typical number of pixels on a camera, meaning that detectors are actually the limiting factor and that most of the information captured by a MO is simply discarded. In the past few years, this mismatch was further increased by the design of high numerical aperture (NA) MO with moderate magnification \cite{singh2015comparison}. In practice, these high performance objectives offer an excellent light collection and spatial resolution but they are underemployed as only a fraction of the FOV can be effectively detected by a camera (see figure \ref{fig1}.a). 

As numerous biomedical applications require a large FOV, the SBP of a microscope is usually increased by mechanically scanning the sample laterally (Fig. \ref{fig1}b), acquiring images for various positions and finally stitching them. Yet, moving the specimen stage with accuracy is intrinsically a slow process that could furthermore be challenging in the case of bulky specimens or mounts. Furthermore, this can induce problematic sample motion in the case of unfixed specimens in an immersion medium.

In this article, we report on a simple method to significantly increase the FOV of a camera-based microscope. Our technique is inspired by the remote focusing method where a second objective in the detection unit is used to refocus the light from the sample and hence to get an axial scan of the sample without moving it \cite{botcherby2008optical}. Here, we propose to perform a transverse scan of the image of the sample without moving it. Practically, a mirror on a motorized kinematic mount is added in the detection part of the microscope. Depending on the tilt applied on the mirror, the camera captures different areas of the FOV detected by the MO. This apparatus allows sequential acquisition of diffraction limited images corresponding to different areas of the microscope FOV. By stitching them, we nearly cover the full microscope FOV resulting in a throughput of around 16 MPixels. In this work, we implement this strategy on both wide-field microscopy and full-field optical coherence tomography (FF-OCT) and we demonstrate the benefits of the technique to image large 3D biological samples.

\section{Results}
\subsection{Principle of remote scanning}

\begin{figure*}
    \centering
    \includegraphics[width=0.85\textwidth]{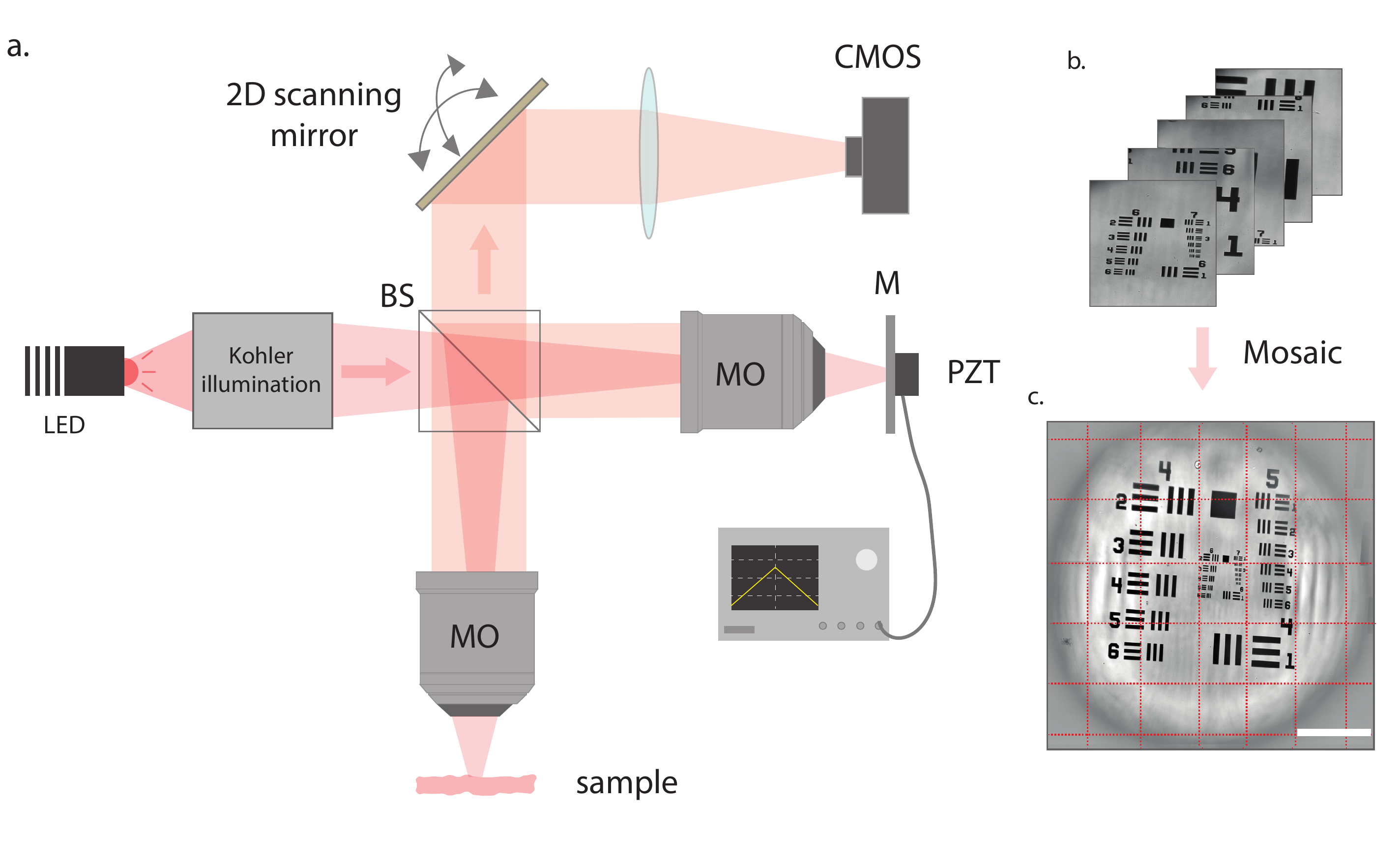}
    \caption{Remote scanning apparatus. (a) Optical setup. MO, microscope objective; BS, beam splitter; M, mirror;  PZT, piezoelectric transducer. A motorized mirror placed in the Fourier space in the detection unit allows to scan different regions in the accessible FOV. (b,c) A stack of images is rearranged as a mosaic to obtain a large FOV. Scale bar, 500 $\mu$m}
    \label{fig2}
\end{figure*}

 In this work, the remote scanning approach is applied on an apparatus that combines both a wide-field mode and full-field optical coherence tomography mode (FF-OCT) \cite{beaurepaire1998full,dubois2002high}. Figure \ref{fig2}.a shows a schematic of the experimental set-up. It is based on a Michelson interferometer illuminated by a LED through a Kohler illumination and with both arms equipped with identical 10X MO. Light reflected by both the specimen and a silicon mirror placed in the focal plane of the two MOs is then recombined and focused onto a CMOS camera. According to the specifications of the MO, the theoretical accessible FOV is 2.65 $\times $2.65 $\text{mm}^2$. However, in the absence of the remote scanning unit, the microscope FOV is restricted by the detection. With a camera whose number of pixels is 1024x1024, Nyquist sampling fulfillment leads to a FOV of only 490 $\times$ 490  $\mu\text{m}^2$. This is roughly 30 times less smaller than the theoretically accessible FOV. In order to match the detected FOV with the MO FOV, we introduce a remote scanning mechanism in the detection part of the set-up.
 
To do so, we simply add a two-axes motorized mirror mount in the Fourier space of the detection arm of the microscope (see methods). In practice, access to Fourier space in the detection is obtained by using an afocal system of magnification 1 (not shown in Fig. \ref{fig2}.d). Because the scanning system is in the Fourier space, a tilt of the mirror is solely translated into a shift of the image in the camera plane, without introducing any time-delay or image distortion. To cover the full FOV of the MO, a series of two-dimensional tilts is applied to the mirror (Fig. \ref{fig2}c).

\subsection{Optical performances}

\begin{figure}
    \centering
    \includegraphics[width=0.45\textwidth]{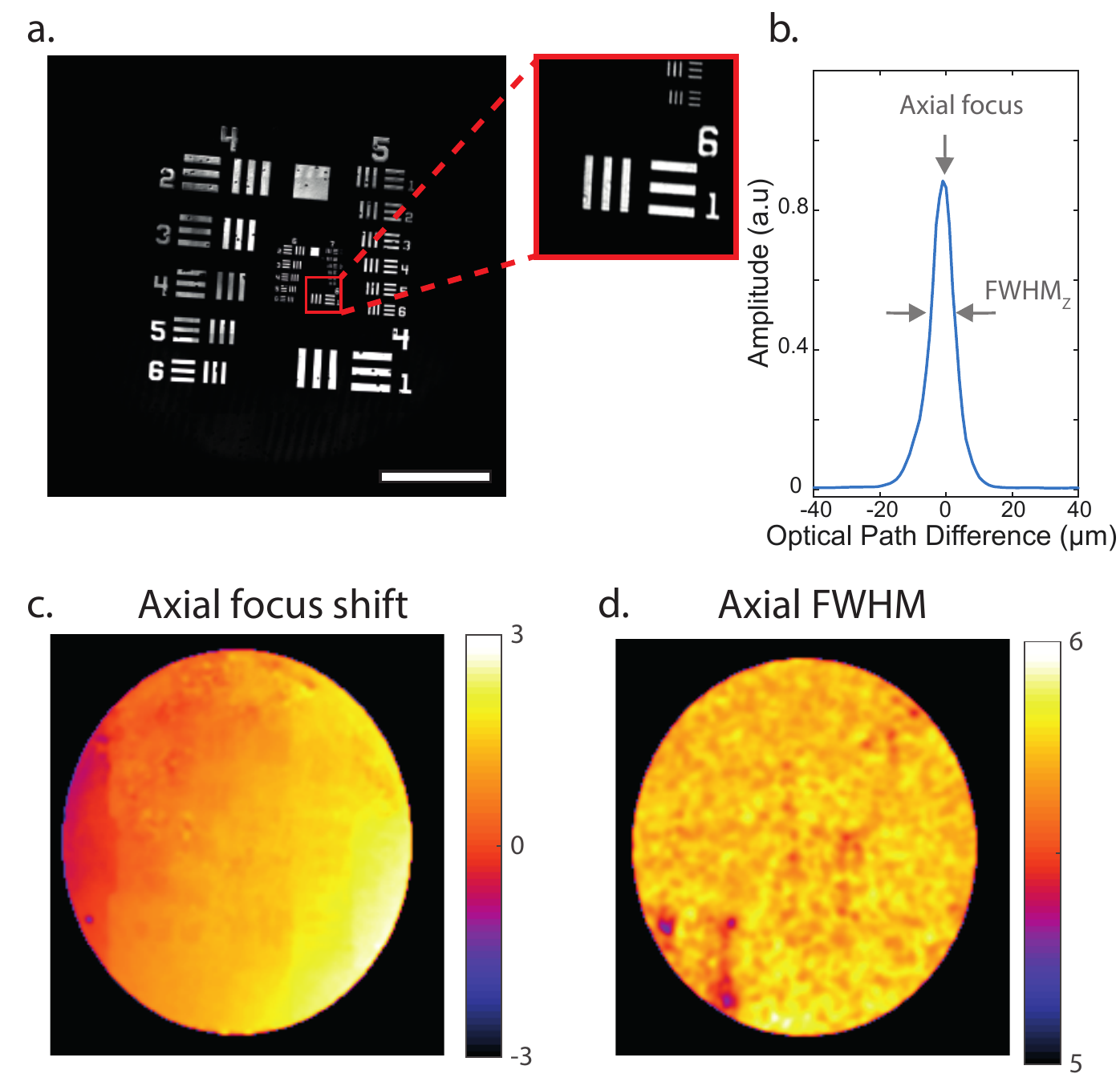}
    \caption{Optical performances of the remote scanning unit applied on FF-OCT. (a) OCT mosaic image obtained by stitching 49 images. (b) Amplitude of the interferometric signal as a function of the optical path difference for blue square in (a). From these curve, the axial focus position and the axial resolution is measured. (c,d) 2D map of the axial focus shift and the axial resolution respectively. Scale bar, 500 $\mu$m. }
    \label{fig3}
\end{figure}

To characterize our optical system in a wide-field microscopy mode, we first imaged a positive USAF 1951 resolution target. By tilting the mirror in the detection unit, 7 $\times$ 7 wide-field images corresponding to different areas of the sample are collected. After stitching these images into a mosaic, a FOV covering 2.2 $\times$ 2.2 $\text{mm}^2$ is obtained with a diffraction limited resolution, here 1.1 $\mu$m. On the edges of the FOV, we observe only a slight loss of signal corresponding to the limits of the field of full light but no transverse resolution degradation (see Supplementary Information). 

Secondly, we assessed the performances of the remote scanning applied to the FF-OCT system. By axially moving the mirror in the reference arm of the interferometer, a z-stack is acquired for each position of the scanning mirror. For each axial position, the electromagnetic field is computed from intensity measurements using a phase shifting interferometry technique (see methods). While the transverse resolution is limited by the numerical aperture of the MO, similarly to the wide-field mode, the axial resolution is given by the coherence length of the light source. Indeed, interferences disappear if the optical path length between the two arms is larger than this physical value. The axial resolution is then estimated as the full width at half maximum (FWHM) of the interferometric signal amplitude (Fig. \ref{fig3}b). From the stacks acquired for the various angles of the scanning unit, the axial resolution is estimated for all the locations of the FOV. As seen in the resolution map shown in figure \ref{fig3}d, our system presents an axial resolution of 5.6 $\mu$m on average. This value matches the coherence length of the LED used as a light source. Notably, no degradation of the axial resolution is observed on the
edges of the FOV.

Finally, the position of the peak in the interferometric signal indicates the axial position of focus. As seen in figure \ref{fig3}c, there is a shift of 6 $\mu$m from the left to the right of the image. We found that this shift is induced by a slight tilt of the silicon mirror in the reference arm and not by the remote scanning system itself. Even though it can be ultimately corrected, note that this value is less than 0.3\% of the FOV. 

Here, we have demonstrated that similar features are obtained for all the images taken at different locations of the FOV. Hence, the remote scanning unit does not alter the nominal optical performances of the microscope, indicating that high-resolution imaging of large specimens is within reach with our approach.

\subsection{Imaging large biological specimen}

\begin{figure*}
    \centering
    \includegraphics[width=1\textwidth]{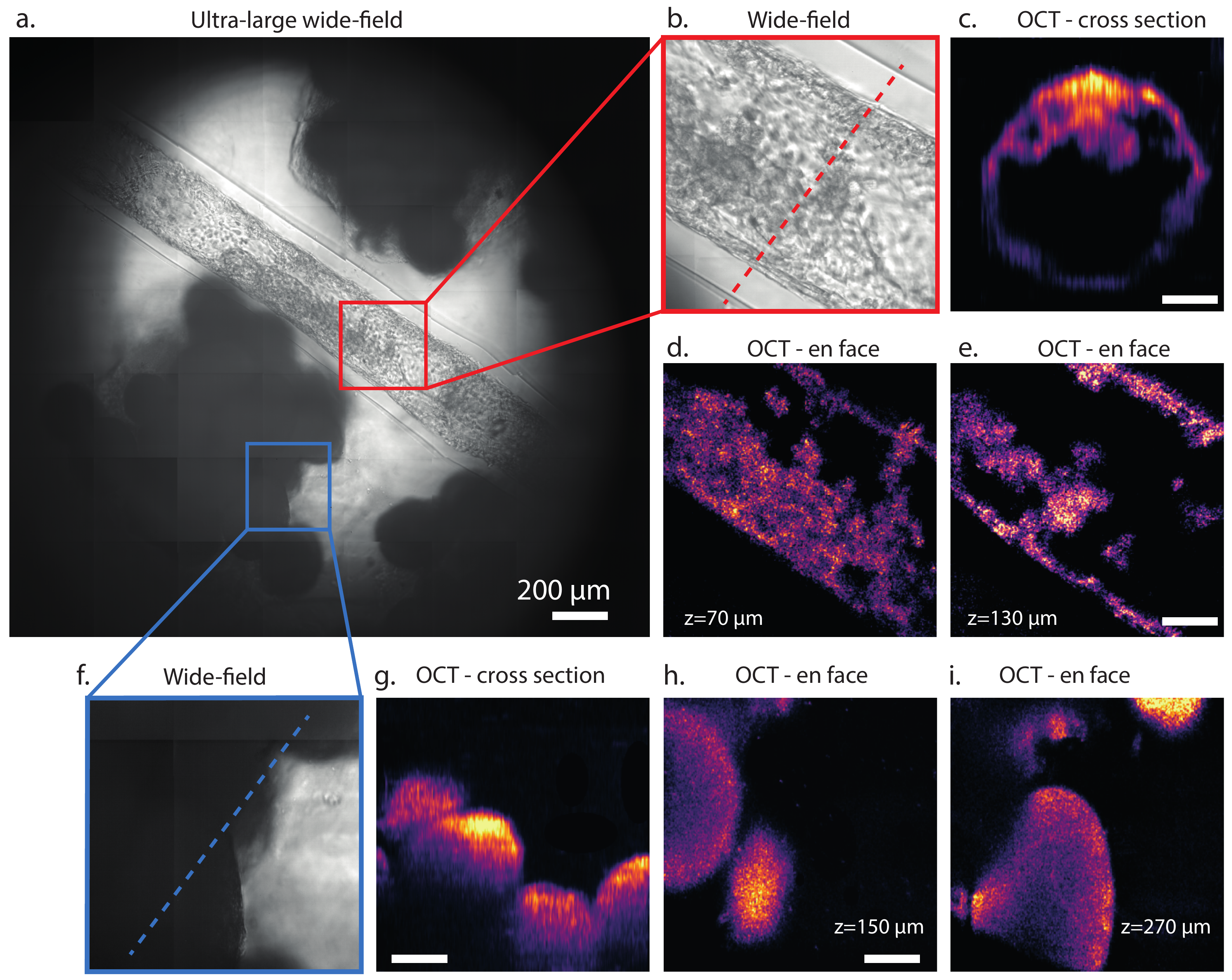}
    \caption{Remote scanning applied to a large sample made of a vessel surrounded by spheroids.(a) Ultra-large FOV in wide-field microscopy  obtained by stitching 7$\times 7$ images. (b) Single wide-field image corresponding to the red square in (a). Cellular information is clearly visible on the vesseloid. (c) Cross section image obtained with full-field OCT. It corresponds to the red dotted line in (b).  (d,e) En face OCT image of the vessel obtained at depth z=70 and z=130 $\mu$m respectively. (f) Wide-field image corresponding to the blue square in (a). (g) Cross section image corresponding to the blue dotted line in (g). (h,i) En face OCT images of the same area at depth z=150 and 270 $\mu$m respectively. Scale bar, 100 $\mu$m. }
    \label{fig4}
\end{figure*}

In the perspective of using our new approach to image a large and challenging biological sample, we applied our technique to an engineered heterogenous cell assembly composed of an endothelial vesseloid (i.e. engineered blood vessel) \cite{andrique2019model} and of encapsulated spheroids of liver cells \cite{alessandri2013cellular}. The overall construct could then be regarded as a prototypal liver tissue architecture, with thick and diffusive micro-tissues in regard with hollowed cell tubes with a lumen that is supposedly conducting blood diffusing molecules and circulating cells. Here, one needs both to visualize how the different building blocks (vesseloid and spheroids) are arranged respectively, and to gain insight into the internal structure of each component.

A 2.2 $\times$ 2.2 $\text{mm}^2$ FOV is first acquired by stitching 7 $\times$ 7 wide-field images for various angles of the scanning unit. With such an extended FOV, we can check that the long vessel is surrounded by packs of spheroids (Fig. \ref{fig4}a). For a single detection angle (a static unitary FOV) we distinguish with a diffraction limited resolution cellular structures without being capable of figuring out whether they are on the surface or inside the vesseloid ( Fig. \ref{fig4}b). However, the mosaic reconstruction yields a more comprehensive picture of the sample since contrast heterogeneities can be detected along the vesseloid. After determining this overall arrangement of the tissue building blocks, we see that not all the regions are equally interesting. To gain efficiency, we decided to focus on the more relevant and optically challenging regions of interest and perform detailed investigation with the FF-OCT mode. This mode is complementary to the wide field mode as it overcome the lack of optical sectioning required for volumetric imaging \cite{badon2017multiple}. 

Here, we focused on a vessel section and an ensemble of packed spheroids (respectively red and blue square on figure \ref{fig4}.a). A z-stack is acquired with a 5 $\mu$m step on each of these areas. These two volumes are obtained without moving the sample, by just adjusting the scanning unit for setting the lateral location and the reference mirror for setting the depth. On the first region, en face OCT images taken at 70 and 130 $\mu$m display different tube widths and inhomogeneous thicknesses (see figure \ref{fig4}.d,e). This complexity is also visible on the cross section image (Fig. \ref{fig4}c). We recognize the characteristic tubular shape of the vessel and thicker structures on the top part. That are the reminiscent shadows of supernumerary cells, that would be washed out upon perfusion. The strengths of FF-OCT are even more striking when imaging the stacks of multicellular spheroids. Indeed, while a contrasted image of a single spheroid is difficult to obtain by wide-field microscopy due to the highly scattering and absorbing properties of the spheroid, the loss of information becomes critical when spheroids are densely packed (Fig. \ref{fig4}f). From en face OCT images taken at 150 and 270 $\mu$m (see figure \ref{fig4}h,i) we see that several spheroids are indeed packed in the dark area of the wide-field image. These spheroids are of different sizes and located at different depths. A cross section image corresponding to the blue dotted line in figure \ref{fig4}f validates this observation (see Fig. \ref{fig4}g). We clearly see the volumetric arrangement of the capsules over 300 $\mu$m axially. Notably, even capsules on top of each other can be imaged.

\section{Discussion}

The remote scanning unit we added on our apparatus provides a large increase of the FOV obtained with both a wide-field microscope and a FF-OCT system. By capturing 7 $\times$ 7 images for different tilt angles in the Fourier plane, we obtained a 2.2 $\times$ 2.2 $\text{mm}^2$ FOV with a diffraction limited resolution, here 1.1 $\mu$m.

Of course, this increase in information requires to make concession in terms of temporal resolution. There is a linear relationship between the number of frames, or acquisition duration, and the amount of data.  Actually, the same trade-off occurs when the sample is moved directly using a motorized stage. Yet, in our approach we move the complexity from the sample plane to the detection unit of the microscope. By taking advantage of the magnification of the system, similar performances as high precision motorized stages are obtained with relatively cost-effective components. In addition, angular scanning is intrinsically much faster than mechanical scanning. In our configuration, it takes approximately 1 minute to acquire the ultra-large FOV in the wide-field mode. Moreover, by replacing the motorized mounts by galvanometric mirrors, this acquisition time can be reduced at least by one order of magnitude. 

As our approach aims to acquire sequentially the total amount of spatial information transmitted by the MO, one could argue that a more straightforward solution would be to increase the number of pixels of the camera. This is true until a certain point. Cameras with higher number of pixels exist but there are usually expensive and even the best ones cannot match the SBP of the most effective MO. Indeed, not all the MO transmits the same amount of information. This quantity depends on the field number (FN), the magnification (Mag) and the NA of a MO as follows \cite{bumstead2018designing}:
\begin{equation}
    SBP  = 4 \times \frac{\left(\frac{FN}{Mag}\right)^2}{\left(\frac{\lambda}{2 NA}\right)^2}
\end{equation}
Using this equation, the SBP for several MO is estimated for $\lambda$=650 nm and given in the table \ref{SBP}. 
\begin{table}
\caption{Spatial bandwidth product for various microscope objectives used in wide-field microscopy and FF-OCT.}
\begin{center}
\begin{tabular}{ |c|c|c|c| } 
\hline
Microscope objective & NA  & SBP (Megapixels) \\
  \hline
Thorlabs TL4X-SAP  X4  & 0.2 & 45 \\ 
Olympus UMPLFLN 10X  & 0.3 & 21 \\ 
Olympus UMPLFLN 20X  & 0.5 & 15 \\ 
Olympus UMPLFLN 40X   & 0.8 & 9  \\
Nikon CFI Super FLuor  X60  & 0.85 & 5 \\ 
  \hline
Nikon N16XLWD-PF  16X  & 0.8 & 97 \\ 
Olympus XLUMPLFLN X20    & 1 & 46 \\ 
Olympus  XLPlan N 25X    & 1.05 & 22 \\ 
  \hline
\end{tabular}
\label{SBP}
\end{center}
\end{table}
From this table, we first see that low magnification MO usually provides a higher SBP than high magnification ones. For instance, using a 4X and 0.2 NA can lead to SBP larger than 45 Mpixels, while a conventional 60X with NA=0.85 has only 5 Mpixels. In the latter case, the benefits of our approach would be very limited as the SBP roughly match the number of pixels of a conventional camera. On the contrary, no current camera can match the SBP of objectives recently developed for multi-photon and light sheet microscopy. In the case of the Nikon N16XLWD-PF, the collection of approximately 100 frames using the remote scanning principle would provide a 2 mm FOV with a 400 nm resolution. 

From this table, the second information is that the SBP of our microscope is still lower than the theoretical SBP of the MO. Indeed, we acquired only 75 $\%$ of the 21 Mpixels. This number is mostly limited by a small amount of clipping in both the illumination and the detection of our system. In addition, field number of optical elements are indications given by the manufacturers and we do not know in which extent theoretical performances can be obtainable.

An other strategy to increase the amount of information collected by a microscope is to use an array of cameras, as proposed in a recent work \cite{Fan2019}. While there is no trade-off between temporal resolution and information with this approach, its complexity and its cost can be an obstacle to its dissemination. An elegant alternative consists in the acquisition of low magnification images obtained for various illumination angles. High resolution images over the entire FOV are then obtained thanks to a stitching operation in the Fourier space \cite{Tian:15,zheng2013wide}. However, this technique called ptychography is mostly used in the transmission mode, and thus non-compatible with thick biological tissues. In addition to its simplicity, our approach offers flexibility as we can tune the number of acquired frames depending on the MO we use and match in each case its SBP. In the case of a non-continuous specimen, there is also the possibility to scan only few regions of interest. This would be similar to a wide-field version of random access scanning microscopy \cite{salome2006ultrafast}.  

Finally, as we were interested in imaging large 3D scattering structures without any label we limited this work on extending the FOV only in the case of wide-field microscopy and FF-OCT. Yet, our approach is not limited to these two imaging techniques and the remote scanning approach can be applied to all camera based or wide-field microscopic techniques. In particular, we believe our approach is of interest for differential interferometric contrast microscopy \cite{shaked2012biomedical}, light sheet microscopy \cite{huisken2004optical} and wide-field fluorescence microscopy \cite{Mohammed2016}. In the latter case, this feature could be of interest to capture neuron activity on large parts of the mouse cortex. 

\section{Conclusion}

In this work, we have introduced a remote scanning mechanism to largely increase the FOV of microscope without sacrificing the lateral resolution and, more importantly, without moving the specimen. By collecting frames for various positions of the scanning unit, we imaged a FOV up to 2.2 $\times$ 2.2 $\text{mm}^2$ with a 1.1 $\mu$m resolution. We demonstrated the benefits of our approach on both wide-field microscopy and FF-OCT. Our approach is particularly valuable when using water immersion objectives.

\section{Methods}

\subsection{Hardware setup}
A broadband and spatially incoherent light source centered at 660 nm (M660L4, Thorlabs) illuminates a Michelson interferometer in a Kohler configuration made of four lenses and two apertures (see SI for details). Light is separated by a beam splitter (CCM1-BS013, Thorlabs) and propagates through two identical MO (UMPLFLN 10X, Olympus) placed in the two arms. The specimen of interest is placed in the focal plane of one of the MO while a silicon mirror is positioned in the second arm. The latter is supported by a piezoelectric transducer (PZT) which modulates the optical path difference between the two interferometric arms. Light reflected in the two arms is then collected by the same MO, reflected by the mirror on the motorized mount (KS1-Z8, Thorlabs), focused by a tube lens (AC508-400-A-ML,f=400 mm, Thorlabs) and imaged by a CMOS camera (MV1-D1024E-160-CL-12, PhotonFocus). Amplitude and phase of the interferometric signal are extracted using 10 consecutives frames acquired during a triangle modulation of the PZT \cite{federici2015wide,thouvenin2017optical}. 
The setup was controlled from a workstation and the experiments were conducted using Matlab.

\subsection{Sample preparation}
The super-assembly of Vesseloids and spheroids was obtained by positioning both tissue blocks within a hydrogel mold (2\% agarose) casted so that a groove receive the Vesseloid and that deep wells can fit a superposition of up to 3 spheroids on top of each other. The spheroids and the Vesseloids were obtained with the Cellular Capsule Technology, used to produce either spherical hollow alginate capsules \cite{alessandri2013cellular} or meter-long tubular capsules \cite{andrique2019model}. In the present study, the cells grown in the spherical capsules were the hepatocarcinoma cell line HuH6. Both spheroids and Vesseloids were fixed overnight in 4\% PFA before being positioned in the agarose mold to prevent tissue degradation.

\subsection{Image processing}
To stitch the different images into the ultra-large FOV, we calibrated the 2D motorized motor using the USAF resolution target. Once the relationship between tilt-angle in the Fourier space and pixel shift in the image plane is known, a simple Matlab script was used to perform this operation. Visualizations were performed using Fiji \cite{schindelin2012fiji}.

\section*{Funding Information} 
This work was supported by the ANR MecaTiss (ANR- 7-CE30-0007-03), Parkington (ANR-17-CE18-0026-02), MecanoAdipo (ANR-17-CE13-0012-01), the charities Ligue contre le Cancer and Canc\'erop\^ole GSO.

\section*{Acknowledgements}
The authors thank M. Caumont for spheroids production, N. Courtois-Allain and F. Saltel for providing us the HUH6 cell line.

\section*{Disclosures}
The authors declare no conflicts of interest.

\bibliography{optica}

\end{document}


\title{Remote scanning for ultra-large field of view in wide-field microscopy and full-field OCT: supplementary material}

\author{Ga\"elle Recher}
\affiliation{LP2N, Laboratoire Photonique Num\'erique et Nanosciences, Univ. Bordeaux, F-33400 Talence, France}
\affiliation{Institut d\textquotesingle Optique Graduate School $\&$ CNRS UMR 5298, F-33400 Talence, France}
\author{Pierre Nassoy}
\affiliation{LP2N, Laboratoire Photonique Num\'erique et Nanosciences, Univ. Bordeaux, F-33400 Talence, France}
\affiliation{Institut d\textquotesingle Optique Graduate School $\&$ CNRS UMR 5298, F-33400 Talence, France}
\author{Amaury Badon*}
\affiliation{LP2N, Laboratoire Photonique Num\'erique et Nanosciences, Univ. Bordeaux, F-33400 Talence, France}
\affiliation{Institut d\textquotesingle Optique Graduate School $\&$ CNRS UMR 5298, F-33400 Talence, France}

\email{amaury.badon@institutoptique.fr}
\date{\today}

\begin{abstract}
This document provides details related to the experimental setup, complex field acquisition and transverse resolution of the wide-field mode.\end{abstract}

\maketitle

\section{Details on experimental setup}

\begin{figure*}
    \centering
    \includegraphics[width=0.85\textwidth]{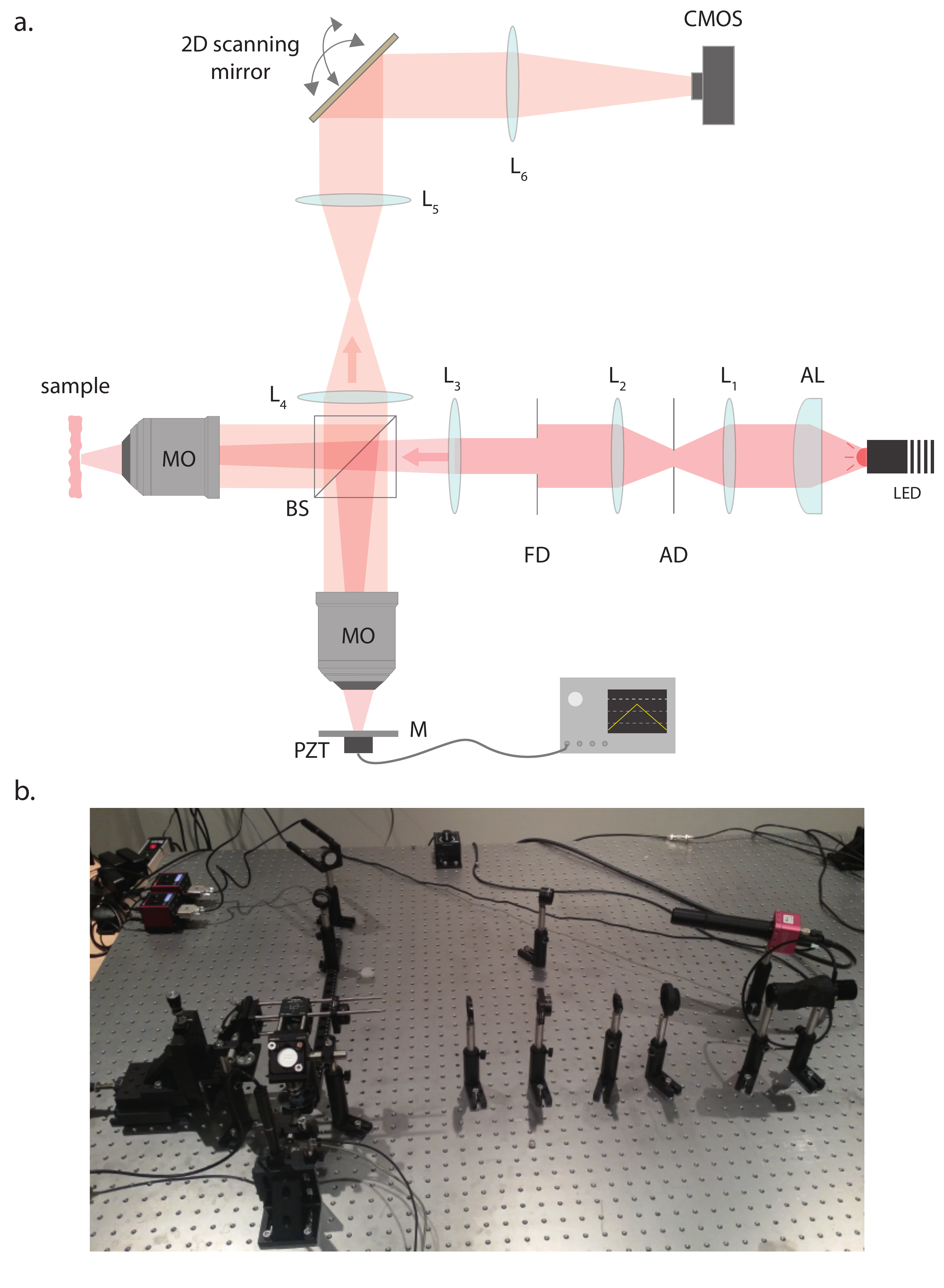}
    \caption{Detailed experimental setup. (a) Schematic of the optical system. FD, field diaphragm; AD, aperture diaphragm; AL, aspheric lens; L, lens; BS, beam splitter; MO, microscope objective; PZT, piezoelectric transducer; M, mirror. (b) Picture of the optical system.}
    \label{figS1}
\end{figure*}

 Our experimental setup consists of a standard full-field optical coherence tomography system with the remote scanning unit added in the detection part (see figure \ref{figS1}.a). Compared to the schematic in the main text, we added the Kohler illumination unit and the afocal system in the detection part. The former is comprised of an aspheric lens (AL2550M-A,Thorlabs), and 3 doublet lenses (AC254-100-A-ML, AC254-100-A-ML and AC254-200-A-ML Thorlabs). A first diaphragm is positioned between $L_1$ and $L_2$ and acts as an aperture diaphragm. It controls the spatial coherence of the illumination. A second diaphragm is placed between $L_2$ and $L_3$ and acts as an aperture diaphragm. It controls the area illuminated by the LED (M660L4, Thorlabs).  In the detection part of the microscope, the afocal system is built with two identical lenses (AC254-150-A-ML, Thorlabs).

To give a physical picture of the setup, we also added a photograph of the optical table (see figure \ref{figS1}.b). In its current configuration, the microscope is made with on the shelf components. In the future, we aim to add our remote scanning unit on existing commercial microscopes.

\section{Acquisition of the electromagnetic field}
 In optical coherence tomography, it is necessary to measure the field from intensity only measurement. In this work, we adopted a triangular 5 phases stepping approach for the interferometric measurement. It consists in modulating the mirror position in triangles with a frequency equals to the tenth of the camera frequency \cite{federici2015wide,thouvenin2017optical}. From a cycle, we extract two measurements of the complex field, one for the ascending part of the cycle and one for the descending part. From 5 intensity images, the amplitude and the phase can be calculated as follows :
 
 \begin{equation}
A= \sqrt{4(I_2-I_4)^2+(I_1-2I_3-I_5)^2}
\end{equation}
and
\begin{equation}
    \phi=-\text{atan} \left( \frac{2(I_2-I_4)}{I_1-2I_3-I_5} \right) 
\end{equation}

\begin{figure}
    \centering
    \includegraphics[width=0.45\textwidth]{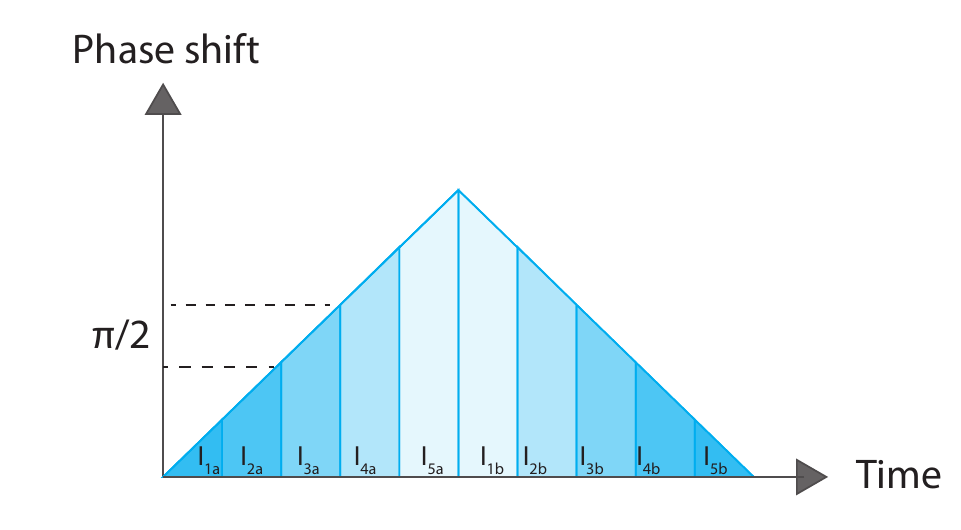}
    \caption{Acquisition scheme. During a modulation cycle, 10 frames are continuously acquired. A combination of 5 frames yields the calculation of the amplitude and phase of the field. }
    \label{figS3}
\end{figure}

\section{Quality of the image depending on the position in the field of view.}
\begin{figure}
    \centering
    \includegraphics[width=0.45\textwidth]{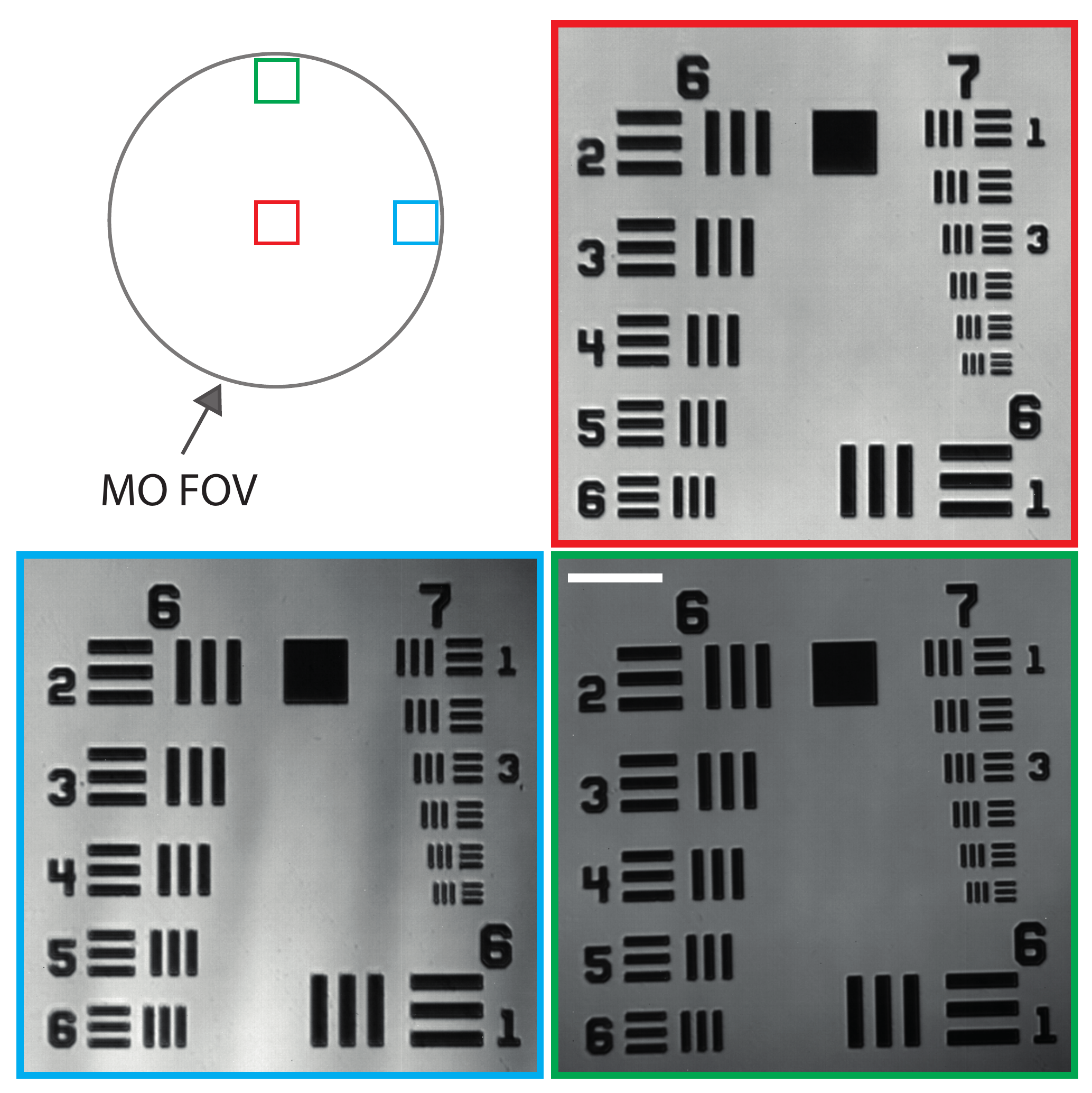}
    \caption{Wide-field image of a USAF resolution target for various locations in the microscope field of view. }
    \label{figS4}
\end{figure}

In this section, we want to check if our remote scanning affects the transverse resolution depending on the location inside the large FOV. To do so, we took wide-field images at the center, on the top and on the left side of the FOV. As seen on figure \ref{figS4}, there is no notable degradation of the transverse resolution on the edges of the FOV. The main effect is a slight loss of signal, as seen on the mosaic image in the main text.

\bibliography{RSOCT}